\documentstyle[prl,aps,twocolumn,epsfig]{revtex}

\begin{document}
\bibliographystyle{prsty}

\title{Mechanisms for Stable Sonoluminescence}
\author{Michael P. Brenner$^1$, Detlef Lohse$^{2,3}$, David
Oxtoby$^{4}$, Todd F. Dupont$^3$}
\address{$^1$Department of Mathematics, MIT, Cambridge, MA 02139 \\
$^2$Fachbereich Physik der Universit\"at Marburg,
Renthof 6, 35032 Marburg Germany \\
$^3$Department of Mathematics, The University of Chicago,
Chicago, IL 60637 \\
$^4$Department of Chemistry and James Franck Institute,
The University of Chicago, Chicago IL 60637}

\date{\today}

\maketitle
\begin{abstract}
A gas bubble trapped in water by an oscillating acoustic field
is expected to either shrink or grow
on a diffusive timescale, depending on the
forcing strength and the bubble size.
At high ambient gas
concentration this has long been observed
in experiments. However,
recent sonoluminescence experiments show that in certain
circumstances when the ambient
gas concentration is low the bubble can be
stable for days. This paper presents mechanisms
leading to stability which predict parameter
dependences in agreement with the sonoluminescence experiments.
\end{abstract}


Recent experiments on
sonoluminescence (SL)
\cite{gai92,bar91,bar94,hil94,bar95,loe93,loe95}
allow detailed studies of the dynamics of a bubble
levitated in a periodically modulated acoustic field.
Besides the light emission itself,
one of the greatest mysteries is how the bubble
can exist in a stable state for many billions of cycles.  Measurements of
the time between successive light flashes show that the
total mass of the bubble remains constant to high accuracy
\cite{gai92,bar91,loe95}.  This result contradicts classical notions
about the dynamics of periodically forced bubbles:  An unforced bubble of
ambient radius $R_0$ dissolves over a diffusive timescale,
$\tau \sim \frac{\rho_0 R_0^2}{D(c_0-c_\infty)}$ \cite{eps50},
where $\rho_0$ is the
ambient gas density in the bubble, $D$ is the diffusion constant of the
gas in the liquid, $c_0$ is the saturated concentration of the gas in
the liquid, and $c_\infty$ is the concentration of gas in the liquid
far from the bubble.
A strongly forced bubble grows
by rectified diffusion, as first discovered by Blake \cite{bla49,ell70}.
This is because
when the
bubble radius is large, the gas pressure in the bubble is small, resulting
in a strong mass flux into the bubble.  Conversely, when the bubble radius
is small there is a strong mass outflux.
Since the diffusive time scale is much larger
than the very short time
the bubble spends at small radii,
gas cannot escape from the bubble during the compression phase and will
be recollected during expansion, so that
the net effect is bubble growth.
At a special value of the ambient radius $R_0^*$
rectified diffusion and normal diffusion exactly balance.  However
the above arguments suggest that
this equilibrium point is {\em unstable};  if the ambient radius is
infinitesimally different from $R_0^*$ the bubble is pushed away from
equilibrium.

The classic papers on rectified diffusion (see e.g.\ Eller
and Crum \cite{ell69,ell70,cru80}) verified the qualitative picture
described above when $c_\infty/c_0 \approx 1$.
Anomalies between theory and experiment do however exist: of
special note is Eller's \cite{ell69} observation of a stable
oscillating bubble persisting over long periods.

There are two controlled parameters in the SL experiments:
the forcing pressure $P_a$ and the gas concentration $c_\infty$.
The key to the discovery of {\it stable}
single bubble SL (whose existence
completely contradicts the above scenario) by Gaitan et al.\
\cite{gai92}
was that (i) $c_\infty/c_0 \ll 1$, and (ii)
$P_a$ must lie between a lower critical pressure
$\approx 1.1$atm and an upper critical
pressure $\approx 1.3$atm.
When $P_a$ is within this window of stability,
the ambient radius remains constant for
billions of cycles,
as evidenced by
the constant phase $\phi$
of the light emission relative to the oscillatory
forcing.  Barber and Putterman \cite{bar91} showed
that the ``jiggle'' in the phase differs by less than
$50$ picoseconds from cycle to cycle.
Outside the window of stability,
$\phi$ (and hence $R_0$)
varies on a diffusive
timescale \cite{bar95}.  The phase grows (implying growth of the ambient
radius), until the bubble becomes parametrically
unstable
\cite{bre95} and microbubbles pinch off.
Experiments \cite{bar95}
show that $\phi$ can oscillate indefinitely on a diffusive
timescale via this mechanism
(diffusive growth followed by pinching off
a microbubble).

The dependence of stable SL on the gas concentration $c_\infty$
is demonstrated by the
UCLA experiments on pure argon bubbles \cite{hil94,bar95}.
For $c_\infty/c_0$ between
approximately $0.06$ and $0.25$,
$\phi$ oscillates on a diffusive timescale
as described above.
At lower argon
concentration $c_\infty/c_0 = 0.004$ however
the phase becomes perfectly
stable
\cite{bar95}, indicating a stable equilibrium.

The goal of this paper is to suggest mechanisms
leading to stabilization:
When
$c_\infty/c_0$ is decreased at
forcing pressures $P_a > 1.1$atm, the classical
unstable equilbrium point $R_0^*$
undergoes an inverse pitchfork bifurcations and actually {\em stabilizes}.
At even higher forcing pressures, there can be {\em several}
stable fixed points, although far from the equilibrium
point small bubbles
shrink and large bubbles grow.
This mechanism is sufficient to explain
the stable equilibria in the SL experiments.  However, high
pressures and temperatures within the bubble cause nondiffusive
effects \cite{loe95} which can also stabilize the bubble.
Both mechanisms
predict parameter dependences
consistent with
SL experiments.
We suggest that the discretization of the
ambient radius predicted by
the diffusive mechanism provides
a clear experimental signature as to which
effect is primarily responsible for stable SL.

We first set up a formalism for studying the stability
of the equilibrium point,
following Fyrillas and Szeri \cite{fyr94} and L\"ofstedt et
al.\ \cite{loe95}.
Let $c(r,t)$ denote the concentration of gas dissolved
in the liquid a distance $r$ from
the center of the bubble.
For $r>R(t)$, where $R(t)$ is the radius of the bubble,
$c$ satisfies a convection diffusion equation
\begin{equation}
\partial_tc + \frac{R^2 \dot{R}}{ r^2} \partial_r c = D\nabla^2 c.
\label{conc}
\end{equation}
The boundary conditions are given by
Henry's law
$c(R,t)= c_0 P(R,t)/P_0$
and by $c(\infty , t) = c_\infty $. The concentration gradient at
the boundary gives the mass loss/gain of the bubble
$\dot{M} =  4 \pi R^2  D \partial_r c|_{R(t)}.$

These equations
determine the growth
of the bubble as a function of time.  There are two crucial
observations:
First, Eller noted that changing
coordinates to
$h=\frac{r^3 - R^3}{ 3}$ and
$\tau=\int^t R^4 dt$
transfers equation (\ref{conc}) to
the simpler form
\begin{equation}
\partial_\tau c =
D \partial_h\biggl(\biggl( 1+\frac{3 h}{ R^3}\biggr)^{4/3} \partial_h c
\biggr) = 0.
\label{bet}
\end{equation}
For the following it is convenient to define the $\tau$-average
of a function $f(t)$ by $\langle f(t)\rangle_\tau = \int f(t) R(t)^4 dt/
\int R(t)^4 dt$.

The second observation \cite{fyr94,loe95} is that
the bubble radius changes over a much
faster timescale than the ambient
radius.
By averaging equation (\ref{bet}) over the fast time scale,
$\partial_h c(\tau)$
can be computed on the slower diffusive timescale.  Then
the dynamics of the ambient radius is given by
\begin{equation}
R_0^2\frac{dR_0}{d\tau} = D \frac{c_\infty - \langle c \rangle_\tau
}{\int_0^\infty
\frac{dh }{ < 1+\frac{3 h }{ R^3}>_\tau
}}.\label{mass2}
\end{equation}

Equilibrium
points satisfy
\begin{equation}
{\langle p\rangle_\tau \over P_0}
= {c_\infty \over c_0}.
\label{equil}
\end{equation}
The equilibrium is stable if the quantity
$\beta=\frac{d\langle p\rangle_\tau}{dR_0}$ is positive.

Now we proceed to analyze this model.  We
calculate numerically
$\langle p\rangle_\tau$ as a function of $R_0$, for different
driving pressures, by $\tau$ averaging
solutions $R(t)$ of the Rayleigh -- Plesset  (RP)
equation.  The RP equation
\cite{loe93,ray17,lau76}
governs the dynamics of an acoustically forced bubble, and is
given by
\begin{eqnarray}
R \ddot R + {3\over 2} \dot R^2  &=&
{1\over \rho_w} \left(p(R,t) - P(t) - P_0 \right)
             \nonumber \\
       &+& {R\over \rho_w c_w} {d\over dt}
\left( p(R,t) - P(t)\right) - 4 \nu {\dot R \over R} - {2\sigma \over
\rho_w R}.
\label{rp}
\end{eqnarray}
We use parameters corresponding to
\cite{bar92,bar94,loe93}
an air bubble in water: the surface tension of the air-water
interface is
$\sigma = 0.073 kg/s^2$, while water has viscosity
$\nu = 10^{-6} m^2/s$, density
$\rho_w= 1000 kg/m^3$, and
speed of sound $c_w=1481 m/s$. The acoustic field is driven via
$P(t) =
P_a \cos (\omega t)$ with
$\omega/ 2\pi = 26.4 kHz$ and external pressure
$P_0= 1$atm.  The pressure inside the bubble
varies adiabatically like
$p(R) \sim (R^3-a^3)^{-1.4}$.
Here $a= R_0/8.73$
is the hard core van der Waals radius.

Figure 1 shows $\langle p \rangle_\tau /P_0$
for several values of $P_a$.
For small $P_a$ , $\langle p \rangle_\tau $ monotonically
decays with $R_0$, signaling a diffusively {\it unstable} equilibrium. For
example,
when $c_\infty/c_0=1$ with a forcing amplitude of
$P_a=0.8$  the unstable equilibrium occurs
at $R_0\approx 5\mu m $.
Note that for large $R_0$ the bubble may become unstable with respect to
shape oscillations \cite{bre95}.

At large $P_a$ however
$\langle p \rangle_\tau $ develops
oscillations as a function of $R_0$, so for a range of
$c_\infty / c_0$ there are several stable equilibrium points.  These
stable equilibria only occur at low $c_\infty/c_0$, immediately
suggesting a reason why diffusively stable
SL only occurs under these conditions.
As an example, see the inset
of figure 1: when $c_{\infty}/c_0 = 10^{-2}$ and $P_a=1.25$atm, there are
stable equilibria (denoted by small
dots in the figure) at $R_0=6.5,6.8,7.1,7.5,8.0$ and $8.5\mu$m.
To further verify the existence of multiple stable equilibria,
we have solved the full equations
(\ref{conc}) and (\ref{rp})
numerically with a standard finite difference scheme \cite{bre96}.
Figure \ref{num}
shows the ambient radius as a function of
time for two different initial conditions with $P_a=1.25$atm.
In each case, the ambient radius saturates towards a constant
($7.1\mu$m and $8.4\mu$m) at long times.

We now outline in detail the predictions of these
calculations for
the SL experiments.
A standard experimental protocol
\cite{gai92,bar94}
is to
slowly increase the
driving pressure $P_a$.
The initial ambient radius depends on
the preparation of the bubble.
At low
pressures, if the bubble size is below the diffusive equilibrium
curve sketched in figure \cite{scetch},
the bubble shrinks.  As the forcing pressure
is increased, there is a critical pressure
where the bubble size becomes greater than $R_0^*$;
the calculations for $c_\infty/c_0=0.25$ indicate
this occurs near $1$atm, in accord with experiments \cite{bar94}.
Above this forcing pressure,
the bubble grows by rectified diffusion.  When the
ambient radius becomes too large the bubble is parametrically unstable;
in experiments the bubble decreases its radius by pinching
off a microbubble (indicated by the downward arrows in figure 2).
As the forcing
pressure is further increased, the bubble size
tracks the parametric instability line \cite{bre95}.
However, at the
pressure $P_a\approx 1.1$atm stable equilibrium
points appear.
A sketch of this inverted pitchfork bifurcation in the $R_0-P_a$ phase
diagram is shown in
figure \ref{scetch}. If the bubble
is attracted to one of the stable points the radius stabilizes,
causing a discontinuous jump in the ambient
radius, as observed by Barber {\it et. al.} \cite{bar94} for air bubbles.
This analysis predicts a similar jump also for argon bubbles.
The
stable state persists for $1.1<P_a<1.3$;
Above $P_a=1.3$
the equilibrium
point destabilizes, and the bubble must return to diffusive
growth followed by microbubble pinching to survive.
For even larger $P_a$ the bubble becomes unstable with respect to shape
oscillations \cite{bre95}.
This entire scenario suggested by
the classical equations
of diffusive dynamics is in good agreement with the
findings from the SL experiments.

The stable equilibrium points are ultimately due to
oscillations in
$\langle p\rangle_\tau$
as a function of $R_0$, which arise from
resonances in the
Rayleigh-Plesset equation.  Oscillations
even occur
in the maximum radius as a function of $R_0$, so
that in some situations adding more gas to a bubble
decreases its maximum size.
A comparison with the Mathieu equation
is
instructive:  If the
eigenfrequency (in the RP equation this depends on $R_0$)
is an integer or half integer fraction of the forcing
frequency, the amplitude of the oscillations
is anomolously large.

Our predictions are in qualitative
agreement with the experiment;
precise quantitative agreement requires accounting for
several neglected effects.  These
include
realistic heat transfer \cite{mik84,loe93} and equations
of state for the gas\cite{mos94}, as well as
shocks within the
bubble \cite{gre93,mos94}, which cause the light emission
itself.

Perhaps more importantly,
the high pressures and temperatures
within the bubble cause complications which influence the
position and stability of the equilibria.  For example,
the RP equation predicts that the minimum pressure inside the
bubble can become as low as $10^{-3}$atm.
However, the presure cannot fall below the equilibrium
vapor pressure of water around $10^{-2}$atm.
Thus, the RP equation can drastically
underestimate the
pressure inside the bubble near its maximum radius.
Calculations  where the pressure inside the bubble is
given by
$max(0.05$ atm,$P(R(t)))$
show that this shifts the position of the equilibrium point  to
larger $R_0$, without affecting the stability.

The increase of the
mass diffusion constant with increasing temperature and
pressure affects both the stability and
the location of the equilibrium point:
When the pressure and the temperature inside the bubble is high,
the diffusion constant
near the bubble wall is
larger than the diffusion constant in the bulk
liquid.
This results in nondiffusive mass ejection when the bubble is small
\cite{loe95}.

We study this increase in the interfacial diffusion constant
with an extremely simple model:
Whenever
the pressure inside the bubble exceeds
a critical pressure $p_{thres}(R)$ we discontinuously increase
the diffusion constant
near the bubble wall
by a factor $f_{thres}$.
The diffusion constant in
the bulk liquid remains constant,
since high pressures and temperatures are
localized near the bubble wall.

Numerical simulations of the full equations \cite{bre96}
with and without
this effect demonstrate that the unstable equilibrium point
can be stabilized by this effect.
The position of the equilibrium point is in general shifted
to larger radii.
Although it is difficult to determine the precise
parameter dependence of this mechanism without a
more accurate model for the diffusion constant, it clearly
only operates at the high pressures where sonoluminescence occurs.

To summarize, we have presented two different mechanisms leading
to stabilization of the ambient
bubble radius.  Contrary to classical intuition,
the diffusive dynamics themselves
have {\em multiple} stable equilibrium points in the SL regime.
The dependence
of the mass diffusion constant on temperature and pressure
also
leads to stabilization.
Which effect dominates the SL experiments?   We suggest that
this issue can be settled experimentally by
determining whether the ambient radius takes on only
a discrete set of values, as predicted
by the purely diffusive mechanism.

\noindent
{\bf Acknowledgements:} We thank S.\ Grossmann,
S. Hilgenfeldt, and L.\ Kadanoff for
helpful discussions.
This work has been supported by the DOE,
the MRSEC Program of the National Science Foundation at the University
of Chicago, and by the DFG through the SFB 185.

\noindent
e-mail: brenner@cs.uchicago.edu, lohse@cs.uchicago.edu

\noindent


\begin{figure}[htb]
\caption[]{
$\langle p \rangle_\tau/P_0 $ as a function of $R_0$ (in $\mu m$) for
$P_a=0.8$atm, $0.9$atm,$1.$atm,$1.1$atm, $1.2$atm and $1.25$atm,
top to bottom.
Equilibrium corresponds to $\langle p \rangle_\tau/P_0 = c_\infty /
c_0$. The equilibrium is diffusively
stable if the slope $\beta=d\langle p \rangle_\tau/d R_0$ is positive.
{\it Inset:}
An enlargement  of $P_a=1.25$atm. The straight line
corresponds to $c_\infty/c_0=10^{-2}$.  The intersection
of the straight line with the curve correspond to equilibrium points.
When $\beta>0$ (the solid dots in the figure) the equilibrium is stable.
}
\label{fig2}
\end{figure}

\begin{figure}[htb]
\caption[]{
The ambient radius $R_0(t)$ as a function of time, for
$P_a=1.25$atm
and $c_\infty /c_0= 0.01$, with two different initial ambient radii.
Each bubble approaches a {\it different} equilibrium ambient radius,
demonstrating that more than one stable equilibrium may exist.
}
\label{num}
\end{figure}

\begin{figure}[htb]
\caption[]{
Sketch of the stability diagram
in the $R_0-P_a$ phase space.
Upon increasing the forcing $P_a$, a window of diffusive
stability develops
through an inverse pitchfork bifurcation.
Further
bifurcations can
occur. At larger $P_a$ the stability window
closes.
Stable and
unstable branches are marked by $s$ and $u$.
The upper curve shows the parametric instability line.
The long arrows sketch growth by rectified
diffusion followed by micro bubble
pinchoff.
}
\label{scetch}
\end{figure}

\end{document}